\documentclass[conference,compsoc]{IEEEtran}
\IEEEoverridecommandlockouts

\usepackage{ifpdf}

\usepackage{amsmath}
\usepackage{amsfonts}

\usepackage{tabularx}
\usepackage{array}
\usepackage{makecell}
\usepackage{multirow}
\usepackage{booktabs}
\usepackage{subcaption}
\usepackage{graphicx}
\usepackage{xurl}
\usepackage{xcolor}
\usepackage{svg}
\usepackage{float}
\usepackage{placeins}

\usepackage{cite}

\usepackage[hidelinks]{hyperref}

\usepackage{algorithmic}

\usepackage{seqsplit}
\usepackage{textcomp}

\renewcommand\arraystretch{1.4}

\def\BibTeX{{\rm B\kern-.05em{\sc i\kern-.025em b}\kern-.08em
    T\kern-.1667em\lower.7ex\hbox{E}\kern-.125emX}}

\begin{document}

\title{Operational Runtime Behavior Mining for Open-Source Supply Chain Security}

\author{
\IEEEauthorblockN{Zhuoran Tan, Ke Xiao, Jeremy Singer, Christos Anagnostopoulos}
\IEEEauthorblockA{
University of Glasgow\\
Glasgow, United Kingdom\\
\{z.tan.1, k.xiao.1, jeremy.singer, Christos.Anagnostopoulos\}@glasgow.ac.uk
}
}

\maketitle

\begin{abstract}
Open-source software (OSS) supply-chain attacks increasingly rely on runtime behavior, while defenders' evidence depends on telemetry granularity. Some monitoring pipelines preserve concrete artifacts such as domains, IP addresses, commands and file paths, whereas others expose only aggregated behavioral summaries. This paper introduces HetHunt, an observability-aware runtime graph hunting framework for software supply-chain threats. HetHunt maps heterogeneous package-execution telemetry into canonical runtime evidence graphs, learns package-level risk using a relational graph encoder, and converts model-facing edge importance into analyst-facing suspiciousness through environmental filtering and relation-aware benign rarity calibration. Rather than treating explanations as confirmed indicators of compromise (IOCs), HetHunt surfaces telemetry-bounded investigation pivots corresponding to IOC-level, TTP(Tactics, Techniques and Procedures)-level, contextual, or telemetry gap. We evaluate HetHunt on OSPTrack and QUT-DV25, covering cross-ecosystem event-level traces and aggregated Python dynamic-analysis telemetry. On QUT-DV25, rarity-calibrated HetHunt achieves strong edge-derived suspiciousness ranking, with 0.876 AUROC, 0.880 AUPRC, and perfect Precision@100. On OSPTrack, network-scoped rarity improves ranking, but noisy event-level telemetry limits reliable low-FPR alerting. These results show that runtime graph explanations must be calibrated to telemetry observability before supporting operational hunting. Overall, HetHunt bridges package-level risk prediction and analyst-facing runtime investigation in software supply-chain security.
\end{abstract}


\section{Introduction}
OSS supply chain attacks frequently execute malicious logic during installation, import, dependency resolution, or post-install scripts~\cite{183985}. Static analysis~\cite{jia_vulnerability_2025, ZHANG2025103973, LU2024112031,10689592} can miss such behavior when code is obfuscated, generated dynamically, or unavailable~\cite{10.1145/3691621.3694950}. Consequently, defenders increasingly rely on runtime telemetry collected from sandboxes, system-call monitors, and package-analysis pipelines~\cite{10.1145/3560835.3564550,10.1145/2555611,7846383}.

However, runtime telemetry is not uniform. Some traces preserve concrete artifacts such as domains, IP addresses, command strings, and process-level execution context~\cite{11025755, ossf_package_analysis}. Other traces expose only aggregated summaries, such as syscall profiles, TCP state-transition counts, file-access buckets, or normalized network entities~\cite{mehedi2025qutdv}. The same malicious behavior may therefore appear as a concrete IOC-level artifact in one telemetry source, as a TTP-level behavioral pattern in another, or may be absent entirely if the behavior is not triggered during execution. Runtime threat hunting is thus \emph{observability-bounded}: the specificity of investigation evidence depends on what the telemetry records.

This creates a mismatch between model-centric evaluation and analyst-facing hunting needs. Existing supply-chain detectors are typically evaluated as package-level malicious/benign classifiers~\cite{mehedi2025qutdv,10.1145/3691620.3695262,mehedi2026edysecdeeplearningbasedexplainable} or plain source-code-level detection~\cite{10884696,ZHANG2025103973} or dependency scanning~\cite{10.1145/3715728}, while graph explainers such as GNNExplainer~\cite{ying_gnnexplainer_2019} and PGExplainer~\cite{luo2020parameterized} identify subgraphs or features that preserve a trained model's prediction. A faithful
model explanation, however, is not necessarily an actionable hunting pivot: analysts need compact, typed runtime interactions that indicate where to inspect, and missing concrete IOCs may reflect abstracted, unrecorded, or untriggered telemetry rather than failed explanation.

We present Heterogeneous-Entity based Hunting (HetHunt), 
an observability-aware runtime graph hunting framework for OSS supply-chain threats. HetHunt maps heterogeneous runtime telemetry into typed canonical evidence graphs, trains a relational graph encoder for package-level risk, and ranks runtime interactions as analyst-facing pivots. A post-hoc hunting layer converts model-facing edge importance into suspiciousness through environmental filtering and relation-aware benign rarity calibration, producing telemetry-bounded pivots rather than confirmed IOC lists.

We evaluate HetHunt on OSPTrack~\cite{11025755}, with 9,758 cross-ecosystem sandbox
executions, and QUT-DV25~\cite{mehedi2025qutdv}, with 14,271 Python dynamic-analysis traces. 
Our evaluation yields four findings. First, package-level classifiers are useful triage upper bounds, but they do not provide analyst-facing runtime pivots. Second, QUT-DV25 shows that stable aggregated telemetry can support strong rarity-calibrated suspiciousness ranking. Third, OSPTrack shows that event-level telemetry preserves concrete artifacts, but its environmental noise limits reliable low-FPR alerting. Fourth, analyst-usefulness annotation confirms that pivot specificity is telemetry-bounded: QUT-DV25 mainly yields TTP-level or contextual evidence, while OSPTrack yields more IOC-level candidates together with more telemetry gaps and environment-driven artifacts.
These findings lead to four contributions:\footnote{Source data are released for reproducibility at \url{https://anonymous.4open.science/r/DDGRL-2C32/}.}
\begin{itemize}
    \item \textbf{Observability-bounded hunting formulation.} We formulate runtime OSS supply-chain hunting as a graph-based problem whose outputs are bounded by telemetry granularity, distinguishing IOC-level, TTP-level, contextual, and telemetry-gap evidence.

    \item \textbf{Canonical runtime evidence graphs.} We design a compact typed graph schema that maps heterogeneous package-execution telemetry into a shared representation over package, process, file, network, command, and syscall entities.

    \item \textbf{HetHunt suspiciousness framework.} We introduce an end-to-end hunting pipeline that separates package risk, model-facing edge importance, and analyst-facing suspiciousness, using environmental filtering and relation-aware benign rarity calibration to surface runtime investigation pivots.

    \item \textbf{Hunting-oriented evaluation.} 
    We evaluate HetHunt using package-level detection, edge-derived ranking, and analyst-usefulness annotation, showing how telemetry granularity shapes pivot specificity, ranking reliability, and deployment utility.
\end{itemize}

\section{Motivation and Problem Framing}

\subsection{Supply-Chain Threat Runtime Evidence}

Runtime hunting requires more than package-level triage: prior supply-chain detectors commonly report package-level maliciousness or risk scores, but such scores do not localize suspicious behavior within the execution trace~\cite{mehedi2025qutdv,mehedi2026edysecdeeplearningbasedexplainable}. Analysts need compact runtime pivots---such as processes, commands, file accesses, or network endpoints---to guide follow-up inspection and threat-intelligence correlation~\cite{9345788,9672347}.

We use \emph{runtime pivot} to refer to a typed interaction in the execution trace that directs analyst attention. For example, such pivots may correspond to a process opening a network connection, spawning a command-line execution, or reading from or writing to a file. Thus, after package-level triage, the central question is not merely whether a package appears malicious, but which observed runtime interactions can guide analyst inspection.

Edge-level explanations provide one possible bridge from package-level prediction to runtime evidence, but their importance scores remain model-facing: they indicate which interactions help preserve the classifier’s decision, not whether those interactions are operationally suspicious or sufficiently specific under the available telemetry. HetHunt therefore separates package-level risk, model-facing edge importance, and analyst-facing suspiciousness.

\subsection{Observability-Bounded Hunting Evidence}

Runtime hunting evidence is bounded by the telemetry source. Event-level traces may preserve domains, IP addresses, file paths, commands, and process context, whereas aggregated summaries may expose only syscall counts, file-access buckets, TCP state transitions, or normalized endpoints. The same behavior may therefore appear as a concrete IOC, an abstract behavioral pattern, or be absent if the relevant code path is not triggered.

\begin{table}[t]
\centering
\scriptsize
\caption{Telemetry-bounded hunting evidence levels.}
\label{tab:evidence-levels}
\setlength{\tabcolsep}{3pt}
\renewcommand{\arraystretch}{1.08}
\begin{tabularx}{\columnwidth}{@{}l>{\raggedright\arraybackslash}X>{\raggedright\arraybackslash}X@{}}
\toprule
Level & Definition & Example \\
\midrule
IOC-level &
Concrete artifact preserved &
IP, domain, URL, path, command \\

TTP-level &
Behavior preserved, artifact abstracted &
install-time network connection \\

Contextual &
Execution region localized &
package-manager invocation \\

Telemetry gap &
Artifact absent or untriggered &
no payload URL recorded \\
\bottomrule
\end{tabularx}
\vspace{-0.5em}
\end{table}

We call this property \emph{observability-bounded hunting}: the specificity of a surfaced pivot is limited by what runtime telemetry preserves. Table~\ref{tab:evidence-levels} summarizes the evidence levels used in this paper, which describe pivot specificity rather than package severity. This framing lets us compare datasets without assuming that all telemetry sources support concrete IOCs. In our setting, QUT-DV25 often supports TTP-level or contextual pivots through stable aggregated summaries, whereas OSPTrack preserves concrete artifacts such as hostnames, IPs, paths, and commands but includes more environment-driven noise.

\section{Related Work}

We summarize related work across software supply-chain detection, runtime threat hunting, and graph learning/explanation.

\subsection{Software Supply Chain Attack Detection}

Software supply-chain defenses commonly use static analysis, dynamic execution, or differential comparison~\cite{10.1145/3714464}. Static and source-code-centric methods leverage program structure, property graphs, program slicing, or neural representations to detect vulnerable or malicious code patterns~\cite{ZHANG2025103973,10884696}. These methods are effective when source code is available and malicious behavior is statically visible, but are less suitable for obfuscated, generated, environment-dependent, or closed-source components. Dynamic and differential approaches observe
package behavior or compare package versions to detect suspicious updates~\cite{10.1145/3691620.3695262,froh_differential_2023,10.1145/3560835.3564550}. However, they often rely on paired baselines or focus primarily on package-level decisions. HetHunt instead uses sandboxed runtime behavior as an operational data source and asks which observed interactions can serve as hunting pivots after triage.

\subsection{Runtime Hunting and Dynamic Analysis}

Threat hunting systems rely on runtime evidence such as process activity, file access, command execution, DNS queries, and network connections. Runtime application self-protection and interactive application security testing similarly exploit execution-time observations, but primarily for application-level protection or vulnerability discovery rather than package-execution hunting~\cite{7846383,8901378}. Prior work on threat hunting and cyber threat intelligence emphasizes mapping low-level telemetry to investigation artifacts such as IOCs, TTPs, and contextual execution evidence~\cite{9345788,10216378,9458828}; operational platforms(e.g., Velociraptor and Elastic pipelines) illustrate the practical value of endpoint telemetry for analyst investigation~\cite{harwood_velociraptor_2023,9672347}. HetHunt differs by applying runtime hunting to sandboxed package-execution telemetry, where the same canonical graph must support both concrete artifacts and aggregated behavioral evidence.

\subsection{Graph Learning and Explanation}

Graph representations are widely used in security because they encode relationships among code elements, vulnerabilities, processes, files, and network endpoints. Prior work has shown the value of structured graph representations for vulnerability and security analysis~\cite{yin_knowledge-driven_2023},
while R-GCN, GAT, HAN, and HGT provide mechanisms for learning over
typed or attention-weighted graph structures~\cite{10.1007/978-3-319-93417-4_38,velickovic2018graph,10.1145/3308558.3313562,10.1145/3366423.3380027}.
Graph explanation methods such as GNNExplainer and PGExplainer identify nodes, edges, features, or subgraphs that preserve a model's prediction ~\cite{ying_gnnexplainer_2019,10.5555/3495724.3497370}. These methods are useful starting points, but model explanation is not equivalent to threat hunting: a faithful explanation may surface common package-manager commands, registry accesses, temporary files, or system-file reads that are predictive but not actionable. 
HetHunt builds on these techniques, but treats graph explanation
as an intermediate signal that must be calibrated before it can serve
as analyst-facing runtime hunting evidence.

\section{Methodology}
Our framework follows a three-stage pipeline (Section~\ref{sec:hethunt framework}). 
First, raw runtime traces from heterogeneous monitoring sources are mapped into canonical multi-hop heterogeneous graphs. 
Second, a type-aware relational encoder, R-GCN, learns package-level risk from these graphs. Finally, a learned edge-importance module highlights runtime interactions most relevant to the classification decision for downstream investigation.
\subsection{Datasets}

We evaluate HetHunt on two runtime datasets with complementary telemetry granularity. OSPTrack provides cross-ecosystem event-level sandbox traces, while QUT-DV25 provides deeper but more aggregated dynamic-analysis summaries. This allows us to test whether the canonical graph ontology supports both concrete runtime artifacts and abstracted behavioral evidence.

\paragraph{OSPTrack (Extended)}
We build on the published OSPTrack runtime-trace dataset~\cite{11025755},
collected using the OpenSSF Package Analysis framework~\cite{ossf_package_analysis}. The framework executes packages in isolated sandboxes during installation and runtime/import phases, recording file I/O, network sockets, executed commands, and DNS requests. We extend the dataset by replaying the same collection pipeline and applying the same labeling criteria, yielding 9,758 instances across npm, PyPI, crates.io, packagist, and RubyGems.
Table~\ref{tab:ecosystem_package_count} summarizes the ecosystem-level
composition.

\begin{table}[ht]
\caption{Ecosystem Package Count by Label}
\vspace{-5pt}
\label{tab:ecosystem_package_count}
\centering
\scriptsize
\begin{tabular}{lccc}
\toprule
\textbf{Ecosystem} & \textbf{Count} & \textbf{Malicious} & \textbf{Eval Role} \\
\midrule
npm       & 5773 & 1128 (19.5\%) & Train + Test \\
pypi      & 2176 & 853 (39.2\%)  & Train + Test \\
rubygems  & 338  & 277 (81.9\%)  & Train + Test \\
\midrule
crates.io & 1206 & 1 (0.1\%)     & Train only (benign) \\
packagist & 265  & 0 (0.0\%)     & Train only (benign) \\
\midrule
\textbf{Total} & \textbf{9758} & \textbf{2259 (23.1\%)} & \\
\bottomrule
\end{tabular}
\end{table}

Because crates.io and packagist contain too few malicious samples for ranking evaluation, they are retained during training to improve benign behavioral coverage but excluded from malicious-sample evaluation metrics. We report OSPTrack ranking and detection results on npm, PyPI, and RubyGems.

\paragraph{QUT-DV25} 

It~\cite{mehedi2025qutdv} contains dynamic-analysis traces from 14,271 Python packages, including 7,127 malicious packages. In contrast to OSPTrack’s event-level telemetry, QUT-DV25 was introduced as an ML-based package-classification dataset built from aggregated behavioral features in tabular form. Its features summarize runtime behavior through counts, categories, and statistics, including syscall-related features, file-access categories, network and TCP state statistics, and dependency-installation information. This makes QUT-DV25 more suitable for stable TTP-level or contextual evidence than for recovering concrete IOC-level pivots such as exact commands, paths, domains, or payload URLs.
We therefore use its tabular feature representation as a strong non-pivot triage baseline, and separately map the available aggregated behavioral items into canonical graph entities and relations for HetHunt.

\subsection{Canonical Graph Ontology and Construction}
\label{sec:canonical graph}
For cross-source analysis, we define a canonical graph ontology and deterministic construction pipeline that maps raw traces from both datasets into heterogeneous graphs with a shared schema.

\subsubsection{Ontology Definition}
Motivated by general IOC~\cite{9345788,10216378,9458828} entities used in threat hunting, we define six node types and eleven edge types that cover the behavioral categories common to both datasets, as shown in Table~\ref{tab:ontology}. 
The ontology is designed to capture package-triggered runtime behavior using a compact, interpretable schema shared across datasets. Node types represent common analyst pivots---package identity, process context, files, network endpoints, commands, and syscalls---while edge types encode how these entities interact during installation or invocation. This design supports both event-rich telemetry and aggregated behavioral summaries
without preserving source-specific logging formats.

\begin{table}[t]
\caption{Canonical Graph Ontology}
\label{tab:ontology}
\centering
\scriptsize
\setlength{\tabcolsep}{3pt}
\begin{tabular}{@{}llllcc@{}}
\toprule
 & \textbf{Type} & \textbf{Direction} & \textbf{Description} & \textbf{OSP} & \textbf{QUT} \\
\midrule
\multirow{6}{*}{\rotatebox[origin=c]{90}{\textsc{Node}}}
 & \texttt{PKG}     & -- & \makecell[l]{Package identity\\(ecosystem, name, version)}       & \checkmark & \checkmark \\
 & \texttt{PROC}    & -- & \makecell[l]{Process instance \\ (runtime execution context)}       & \checkmark & \checkmark \\
 & \texttt{FILE}    & -- & \makecell[l]{File object \\(path or normalized access bucket)}      & \checkmark & \checkmark \\
 & \texttt{NET}     & -- & \makecell[l]{Network endpoint \\(IP, domain, port)}                 & \checkmark & \checkmark \\
 & \texttt{SYSCALL} & -- & System call identifier                              &            & \checkmark \\
 & \texttt{CMD}     & -- & Command string                                      & \checkmark &            \\
\midrule
\multirow{10}{*}{\rotatebox[origin=c]{90}{\textsc{Edge}}}
 & \texttt{LOAD}       & \texttt{PKG}$\!\to\!$\texttt{PROC}     & \makecell[l]{Package loads or\\ initializes a process}        & \checkmark & \checkmark \\
 & \texttt{DEPEND}     & \texttt{PKG}$\!\to\!$\texttt{PKG}      & Package dependency relationship               &            & \checkmark \\
 & \texttt{INVOKE}     & \texttt{PROC}$\!\to\!$\texttt{SYSCALL} & Process invokes system call                   &            & \checkmark \\
 & \texttt{EXEC}       & \texttt{PROC}$\!\to\!$\texttt{CMD}     & Process executes command string               & \checkmark & \checkmark \\
 & \texttt{SPAWN}      & \texttt{PROC}$\!\to\!$\texttt{PROC}    & Process spawns child process                  & \checkmark & \checkmark \\
 & \texttt{READ}       & \texttt{PROC}$\!\to\!$\texttt{FILE}    & File read access                              & \checkmark & \checkmark \\
 & \texttt{WRITE}      & \texttt{PROC}$\!\to\!$\texttt{FILE}    & File write access                             & \checkmark & \checkmark \\
 & \texttt{DELETE}     & \texttt{PROC}$\!\to\!$\texttt{FILE}    & File deletion                                 & \checkmark &            \\
 & \texttt{CONNECT}    & \texttt{PROC}$\!\to\!$\texttt{NET}     & Network connection (socket-level)             & \checkmark & \checkmark \\
 & \texttt{DNS\_QUERY} & \texttt{PROC}$\!\to\!$\texttt{NET}     & \makecell[l]{DNS query \\(A/AAAA/CNAME lookup)}               & \checkmark &            \\
 & \texttt{RESOLVE}    & \texttt{PROC}$\!\to\!$\texttt{NET}     & DNS resolution result                         & \checkmark &            \\
\bottomrule
\end{tabular}
\vspace{2pt}
\parbox{\columnwidth}{\footnotesize
Node and edge types not observed in a dataset yield no instances and require no special handling in downstream graph construction or encoding.}
\end{table}

\begin{figure}[htbp]
    \centering
    \captionsetup{justification=centering}
    \includegraphics[width=0.5\textwidth]{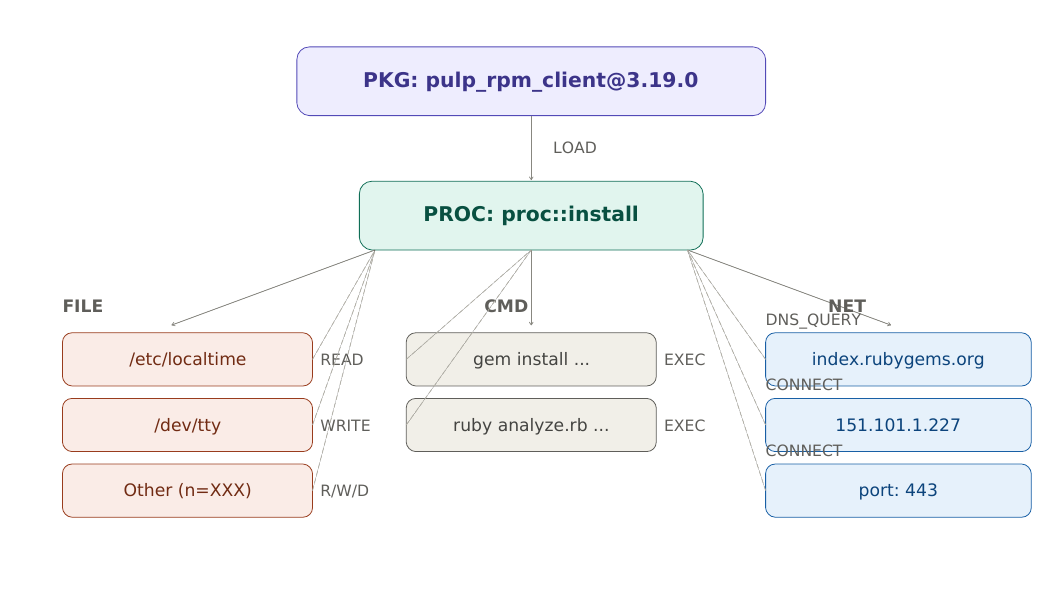}
    \caption{Record-to-Graph Example}
    \label{fig: worked graph}
\end{figure}

This ontology deliberately introduces an intermediate \textit{PROC} layer between \textit{PKG} and behavioral leaf nodes, forming multi-hop evidence paths such as \texttt{PKG}$\!\to\!$\texttt{PROC}$\!\to\!$\texttt{FILE/NET/SYSCALL/CMD}. Figure~\ref{fig: worked graph} shows how this design links package execution to file, command, and network artifacts through process context, allowing HetHunt to distinguish behaviors such as installation-time file writes and import-time network connections within the same package. This relation-specific structure supports downstream message passing, while the sensitivity of relation scope and calibration choices is analyzed in Appendix~\ref{app:sensitivity}.

\subsubsection{Dataset-Specific Parsing}

Each dataset is parsed into canonical edge events following the shared schema in Table~\ref{tab:ontology}. OSPTrack contributes event-level FILE, NET, CMD, and DNS interactions, while QUT-DV25 contributes aggregated file-access, syscall, dependency, command-pattern, and TCP-summary interactions. Dataset-specific parsing details are provided in Appendix~\ref{app:graph-implementation}.

\subsubsection{Heterogeneous Graph Construction and Homogeneous Conversion}
\label{sec:graph construction}

For training and explanation, each package graph is represented as
$G=(V,E,\phi,\psi)$, where $\phi$ and $\psi$ preserve node and relation types. 
This typed edge-list representation provides a compact implementation view: it batches heterogeneous runtime interactions in a single edge set while retaining relation identifiers for type-aware message passing. Nodes are initialized with learnable type embeddings, and R-GCN applies relation-specific
transformations according to $\psi(e)$. The same edge-list view also allows the explainer to assign masks over one shared set of runtime interactions, rather than maintaining separate masks for
each relation-specific tensor. Additional implementation details are provided in Appendix~\ref{app:graph-implementation}.

\subsection{HetHunt Framework}
\label{sec:hethunt framework}

HetHunt is designed to preserve typed runtime semantics, separate risk prediction from edge explanation, calibrate suspiciousness, and interpret pivots under telemetry observability.

\begin{figure*}[htbp]
    \centering
    \footnotesize
    \captionsetup{justification=centering}
    \includegraphics[scale=0.34]{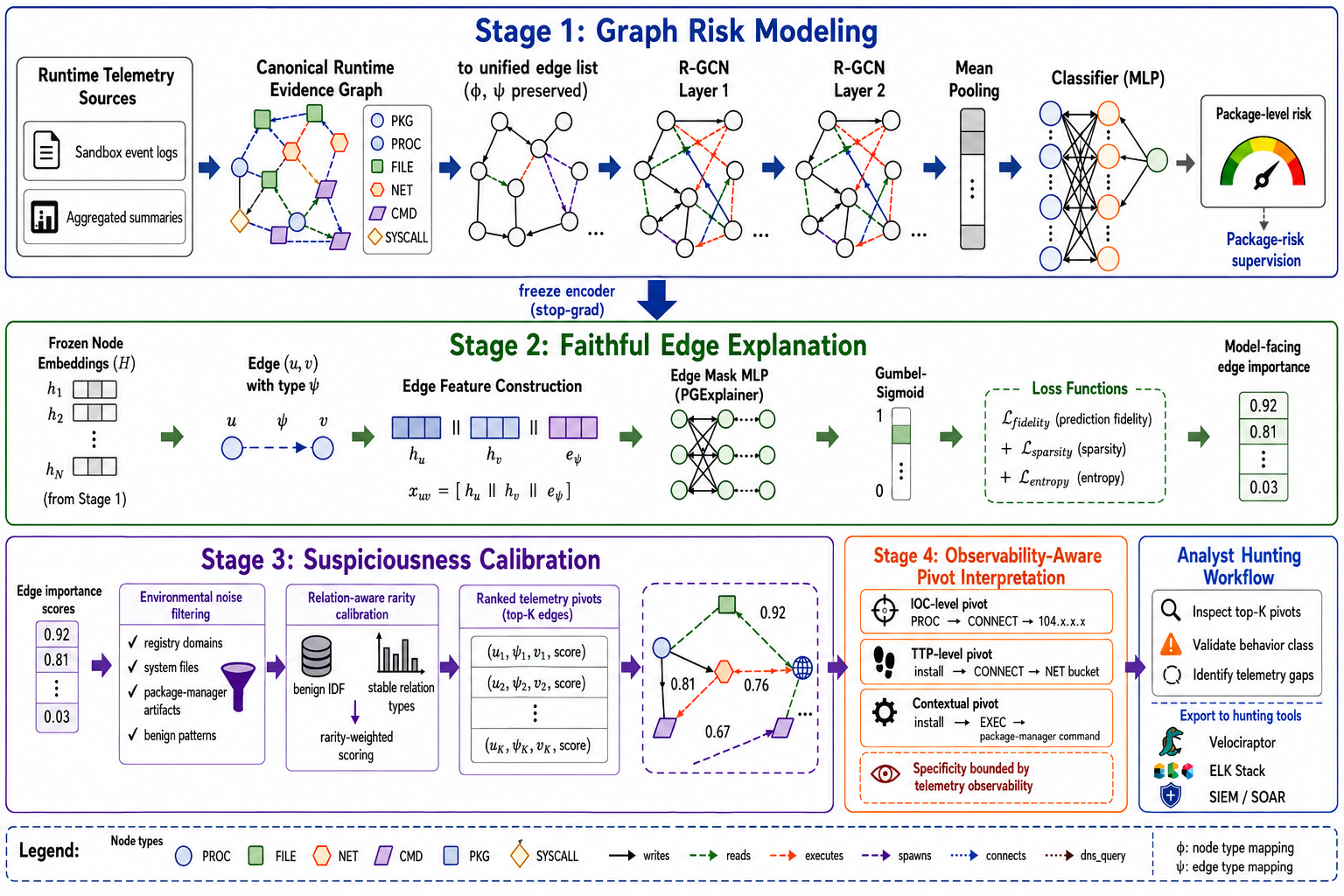}
    \caption{HetHunt Framework with Workflow}
    \label{fig:hethunt framework}
\end{figure*}

As shown in Figure~\ref{fig:hethunt framework}, HetHunt consists of two
decoupled components: a graph encoder that learns node representations from the behavioral graph structure, and an edge importance module that identifies which runtime interactions are most relevant for investigation in threat hunting (Velociraptor~\cite{harwood_velociraptor_2023}, ELK Stack~\cite{9672347}, SIEM/SOAR~\cite{10216378}). 

\subsubsection{Graph Encoder}
\label{sec:graph encoder}

HetHunt uses R-GCN as the default hunting backbone because the pivot-producing stage requires not only package-level risk prediction, but also a stable relation-indexed interface for edge masking, rarity calibration, and analyst-facing pivot reporting. 
The graph encoder operates on the typed edge-list representation defined in Section~\ref{sec:graph construction}, where each edge retains its canonical relation type. 
We stack two RGCNConv layers~\footnote{\seqsplit{https://pytorch-geometric.readthedocs.io/en/stable/generated/torch\_geometric.nn.conv.RGCNConv.html}} with basis decomposition, each followed by ReLU activation and dropout. 
For package-level risk prediction, we obtain a graph-level representation by mean-pooling node embeddings and passing the pooled representation through a linear classifier.

\textbf{Encoder choice and alternatives.} 
We additionally evaluate GAT, HAN, and HGT as package-classification backbones under the same canonical graph construction and feature setup. 
These alternatives test whether different graph encoders improve package-level risk prediction when lightweight canonical node features are available. 
HGT achieves the strongest package-level classification performance (Table~\ref{tab:qut_three_level}), but the final HetHunt pipeline uses R-GCN because R-GCN gives a better operational hunting trade-off: it provides competitive package-risk signal, slightly stronger rarity-calibrated hunting ranking than HGT, lower inference latency than HGT, and a direct typed-edge interface for mapping edge scores back to runtime pivots.

All backbones share the same graph construction pipeline, train/test splits, and evaluation protocol, ensuring that performance differences reflect encoder capacity rather than data preparation artifacts.

\subsubsection{Edge Importance Learning Module}
\label{sec:edge importance}

The explanatory component of HetHunt is a learnable edge-mask module, inspired by PGExplainer~\cite{10.5555/3495724.3497370}, that produces per-edge importance scores identifying which runtime interactions are most influential to the model's prediction. Unlike attention-based ranking (where importance is read from attention weights that serve double duty as message-passing coefficients), the edge-mask module is trained for explanation fidelity and sparsity.

\paragraph{Architecture}
 Given node embeddings $h_{u}, h_{v}$ from the frozen graph encoder for an edge $e=(u, v) \in E$, the module computes an edge score via a small multilayer perceptron(MLP):
\begin{equation}
s_e = \text{MLP}([\mathbf{h}_u \| \mathbf{h}_v \| \mathbf{e}_{\psi(e)}])
\end{equation}
where $\mathbf{e}_{\psi(e)}$ is a learnable embedding for the edge's relation type and $\|$ denotes concatenation. During training, we pass the score through a Gumbel-Sigmoid~\cite{geng2020does} sampler to approximate discrete edge selection in a differentiable manner. This lets the explainer optimize fidelity and sparsity objectives over edge masks with standard backpropagation, while encouraging the learned mask to behave like a compact subgraph selection. The resulting soft mask $m_e \in [0,1]$ serves as the edge-importance score.

\paragraph{Training objective}
The edge-mask module is trained after the backbone encoder is frozen, using a composite loss:
\begin{equation}
\label{eq:edge_mask_training_objective}
L_{\text{total}} = \lambda_{\text{fid}} \cdot L_{\text{fidelity}} + \lambda_{\text{sp}} \cdot L_{\text{sparsity}} + \lambda_{\text{ent}} \cdot L_{\text{entropy}}
\end{equation}
We set $\lambda_{\mathrm{fid}}=1$ and use $\lambda_{\mathrm{sp}}=1.0$ and $\lambda_{\mathrm{ent}}=0.01$ by default. The fidelity coefficient is fixed because preserving the backbone prediction is the primary
objective. The sparsity coefficient controls how strongly the soft mask is regularized toward the target keep ratio, while the entropy coefficient controls the strength of mask sharpening.

Specifically, $\mathcal{L}_{\mathrm{fidelity}}$ preserves the backbone model's predictive distribution under the masked subgraph, $\mathcal{L}_{\mathrm{sparsity}}$ penalizes deviations between the average mask value and a target explanation density, and $\mathcal{L}_{\mathrm{entropy}}$ is applied to the relaxed edge-mask probabilities. For each edge mask $m_e \in (0,1)$, the entropy term penalizes high-entropy values near $0.5$, encouraging more decisive selections close to either retained or suppressed.

\paragraph{Inference output}
At inference time, the explainer assigns every graph edge a continuous importance score, measuring how strongly that runtime interaction contributes to the model’s prediction. These scores capture model influence, not investigative priority: a highly influential edge may still reflect common benign behavior. We therefore apply post-hoc filtering and rarity calibration to convert edge importance into analyst-facing suspiciousness.

\subsubsection{From Importance to Hunting Suspiciousness}

Raw edge-importance scores are not sufficient for threat hunting. In our
inspection of explanation outputs, highly ranked edges often corresponded to
environment-driven or frequently recurring behaviors rather than
package-specific suspicious activity. In supply-chain installation
environments, these behaviors include registry access, system-library reads,
temporary-file management, environment-probing commands, and
analysis-framework artifacts. If used directly, raw importance scores may
therefore surface shared background behavior rather than actionable
package-specific investigation pivots.

To transform model-derived importance into analyst-facing suspiciousness,
HetHunt applies two post-processing stages: (1) rule-based noise filtering to
suppress environmental artifacts and redundant edges, and (2) benign rarity
calibration to prioritize important edges that are also unusual relative to
benign packages.

\paragraph{(1) Rule-based noise filtering}

Before ranking, we remove or suppress known sources of environmental noise. The exact rules are dataset-aware, but the overall strategy is consistent across datasets:
\begin{itemize}
    \item \textbf{Network allowlisting.} Remove common registries and benign services.
    \item \textbf{System-file suppression.} Suppress system paths, devices, pipes, and sockets.
    \item \textbf{Command filtering.} Remove routine probing and analysis-framework commands.
    \item \textbf{Structural filtering.} Drop or demote ubiquitous LOAD-style edges.
    \item \textbf{Diversity control.} Collapse duplicate destinations and cap repeated edge types.
\end{itemize}

These filters do not define maliciousness. Instead, they remove high-frequency
background behavior so that the remaining candidates better reflect
package-specific investigation signals.

\paragraph{(2) Benign rarity calibration}

After filtering, we calibrate edge importance against the benign population.
An edge should be ranked higher if it is both influential to the model's
prediction and uncommon among benign packages.

For each retained edge $e$, we compute a rarity term from the document
frequency~\cite{ramos2003using} of its typed destination in the benign
training set. Specifically, rarity is estimated over
$(\psi(e), \mathrm{dest}(e))$ pairs rather than destination tokens alone, so
that frequency is calibrated with respect to the relation under which an
entity is observed. We then define the edge-level suspicious score as
\[
s(e) =
\mathrm{importance}(e)
\times
\bigl(1+\alpha \cdot \mathrm{rarity}(e)\bigr),
\]
where $\mathrm{importance}(e)$ is the explainer score and
$\mathrm{rarity}(e)$ is computed from the benign reference population. Higher
values of $s(e)$ indicate edges that are both influential to the model and
unusual relative to benign behavior.

Rarity calibration need not be applied uniformly to all relation types. For
datasets with stable and semantically reusable entities, rarity can be
computed over a broad set of edge types. When relation types are dominated by
long-tailed or weakly normalized destination keys, however, applying rarity
indiscriminately may amplify benign outliers. We therefore allow rarity to be
restricted to selected relation types. In our OSPTrack experiments, the most
stable setting applies rarity only to network-related relations
(\texttt{CONNECT}, \texttt{DNS\_QUERY}, and \texttt{RESOLVE}), whereas applying
rarity to all relation types is less reliable due to long-tailed
\texttt{FILE} and \texttt{CMD} entities. This restriction is applied during
edge scoring before graph-level aggregation.

\paragraph{(3) Operational interpretation of edge scores}

HetHunt interprets each retained edge as a typed runtime pivot. Relation types are preserved throughout graph learning, edge-importance estimation, and rarity calibration because suspiciousness is relation-dependent. For example, a file reached through \textsc{Read} may indicate metadata inspection, whereas the same file reached through \textsc{Write} or \textsc{Delete} may indicate modification or cleanup. Similarly, \textsc{DnsQuery}, \textsc{Resolve}, and \textsc{Connect} capture different stages of outbound communication.

At inference time, surfaced edges are candidate investigation leads, not
confirmed IOCs. Depending on what the telemetry source preserves, a pivot may
correspond to a concrete IOC-level artifact, a TTP-level behavior, a
contextual execution region, or a telemetry gap. The effect of rarity scope
and strength is analyzed in Appendix~\ref{app:sensitivity}.

\paragraph{(4) Graph-level suspicious score}

Edge-level suspiciousness is used both to surface runtime pivots and to
prioritize whole packages for analyst review. After filtering, deduplication,
rarity calibration, and any optional per-edge-type caps, we sort the retained
edges in descending order of suspiciousness. Let $\mathcal{E}_{K}(G)$ denote
the top-$K$ retained edges for graph $G$. We define the graph-level suspicious
score as
\[
S(G)=\frac{1}{K}\sum_{e \in \mathcal{E}_{K}(G)}s(e),
\]
where $s(e)$ is the calibrated edge-level suspicious score. Packages are then
ranked by $S(G)$, so that larger values indicate graphs with more concentrated
high-suspiciousness evidence among their top surfaced pivots.

The top-$K$ mean avoids diluting suspiciousness with large amounts of routine
background behavior, especially in dense runtime graphs, while still producing
a compact set of edges for analyst inspection.
\subsection{Training Pipeline and Scalability}

HetHunt follows a two-stage train-then-explain pipeline: the graph encoder is first trained for package-level risk, then frozen while the edge-mask module is trained for explanation. Separate encoders are trained for OSPTrack and QUT-DV25 because their telemetry distributions differ. Scalability and implementation details appear in Appendix~\ref{app:latency} and~\ref{app:graph-implementation}.

\section{Evaluation}
\label{sec:evaluation}

We evaluate HetHunt through five research questions:
\begin{itemize}
    \item \textbf{RQ1:} How much package-level triage signal is available when packages are represented using non-pivot features versus graph-based representations?
    \item \textbf{RQ2:} Can model-facing edge-importance scores be translated into analyst-facing hunting pivots?
    \item \textbf{RQ3:} Which relation types provide stable hunting evidence after rarity calibration?
    \item \textbf{RQ4:} How does telemetry granularity limit the specificity of surfaced pivots?
    \item \textbf{RQ5:} What kinds of telemetry-bounded pivots does HetHunt surface in practice?
\end{itemize}

HetHunt is not intended to replace tabular package classifiers. Following the original QUT-DV25 evaluation setting, processed-feature Random Forest (RF) and Gradient Boosting (GB) models achieve near-perfect package-level triage, but they output package scores rather than typed runtime pivots. We therefore treat them as non-pivot triage upper bounds: they measure dataset separability under engineered summaries, but not which process, file, command, syscall, or network interaction an analyst should inspect.

\textbf{Implementation and reproducibility}. 
We implemented HetHunt using PyTorch Geometric~\footnote{https://pytorch-geometric.readthedocs.io/}, with deterministic dataset-specific parsers that emit the canonical edge-event schema described in Section~\ref{sec:canonical graph}. 
For the QUT-DV25 graph backbones, each node is augmented with lightweight canonical features computed from the graph representation itself, including node type, in/out/total degree statistics, relation-neighborhood summaries, a relative graph-size indicator, and hashed node-key features. 
These features provide graph-local distinguishability among nodes of the same type while avoiding the original processed QUT-DV25 tabular feature vectors and raw string embeddings. 
Package-level graph classifiers are trained with class-weighted cross-entropy, using class weights estimated from the training split. 
The edge-mask explainer is trained after freezing the selected backbone, and the hunting score is evaluated after rarity calibration, filtering, and deduplication. 
For reproducibility, all experiments use a fixed random seed of 42 and a fixed train/test split; the test sets contain 2,570 QUT-DV25 graphs and 1,261 OSPTrack graphs. 
Experiments were run on a workstation with one NVIDIA TITAN Xp GPU with 24GB memory, 64GB RAM, and 8 CPU cores.

\subsection{Ranking-Oriented Evaluation Protocol}
\label{sec:evaluation-protocol}
We organize evaluation by hunting capability rather than by model family. Level 1 provides a non-pivot package-level triage upper bound using processed tabular features. Level 2 tests whether canonical graphs retain package-level triage signal under sparse node features. Level 3 is the only pivot-producing setting and is evaluated as a hunting pipeline: it ranks packages using suspiciousness derived from surfaced runtime interactions.

\paragraph{\textbf{Level 1: Traditional ML on processed features}}
This level uses pre-aggregated QUT-DV25 tabular features, including numerical summaries and encoded categorical/list-valued fields, as strong non-graph package-level baselines. We evaluate Decision Tree (DT), RF, GB, and SVM.

\paragraph{\textbf{Level 2: GNN backbones on canonical graphs}}
This level uses the canonical heterogeneous graphs described in Section~\ref{sec:canonical graph} and evaluates whether graph-structured runtime evidence retains package-level triage signal. 
Unlike the processed tabular baselines in Level~1, the GNN backbones do not consume the original QUT-DV25 engineered feature vectors. 
Instead, each node is initialized with lightweight canonical node features derived from the graph itself, including node type, structural degree statistics, relation-neighborhood summaries, relative graph-size indicators, and hashed node-key features. These features make nodes within the same type distinguishable while avoiding raw string-content modeling.
We compare R-GCN, GAT, HAN, and HGT under the same graph construction pipeline, train/test split, and evaluation protocol.

\paragraph{\textbf{Level 3: HetHunt}}
This level combines an R-GCN risk model, PGExplainer edge attribution, and benign-population rarity reranking. Unlike Levels 1 and 2, it produces edge-level pivots in addition to package-level rankings.
Our primary evaluation uses ranking metrics that reflect realistic analyst workflow:
\begin{enumerate}
    \item Precision@K / Recall@K (K=10, 50, 100): measuring how many malicious packages appear in the analyst's top-K inspection list.
    \item AUROC / AUPRC: threshold-invariant ranking quality measures.
\end{enumerate}

We report classification metrics (Accuracy, Precision, Recall, F1) as a sanity check, but emphasize that these are secondary to ranking quality for the hunting use case. Latency results are reported in Appendix~\ref{app:latency}; in brief, inference cost is dominated by graph size and explanation workload.

\subsection{QUT-DV25 Results}

\begin{table*}[!htbp]
\centering
\scriptsize
\caption{Non-pivot triage baselines versus pivot-producing hunting evaluation on QUT-DV25.}
\label{tab:qut_three_level}
\begin{tabular}{llrrrrrrrrr}
\toprule
Level & Model & Acc & Prec & Rec & F1 & AUROC & AUPRC & P@10 & P@50 & P@100 \\
\midrule
\multirow{4}{*}{Level 1}
& DT  & 0.9802 & 0.9777 & 0.9830 & 0.9804 & 0.9801 & 0.9697 & 1.000 & 1.000 & 0.980 \\
& RF  & 0.9907 & 0.9892 & 0.9923 & 0.9908 & 0.9997 & 0.9997 & 1.000 & 1.000 & 1.000 \\
& GB  & 0.9922 & 0.9923 & 0.9923 & 0.9923 & 0.9997 & 0.9997 & 1.000 & 1.000 & 1.000 \\
& SVM & 0.9693 & 0.9684 & 0.9707 & 0.9696 & 0.9955 & 0.9959 & 1.000 & 1.000 & 1.000 \\
\midrule
\multirow{4}{*}{Level 2}
& R-GCN & 0.7265 & 0.7980 & 0.6127 & 0.6931 & 0.8124 & 0.8127 & 1.000 & 0.980 & 0.970 \\
& GAT   & 0.7307 & \textbf{0.8139} & 0.6042 & 0.6935 & 0.8199 & 0.8205 & 1.000 & 0.980 & 0.920 \\
& HAN   & 0.7591 & 0.7982 & 0.6991 & 0.7454 & 0.8554 & 0.8640 & 1.000 & 0.980 & \textbf{0.990} \\
& HGT   & \textbf{0.7895} & 0.7798 & \textbf{0.8117} & \textbf{0.7955} & \textbf{0.8859} & \textbf{0.8950} & 1.000 & \textbf{1.000} & \textbf{0.990} \\
\midrule
\multirow{2}{*}{Level 3}
& HetHunt (filtered)    & -- & -- & -- & -- & 0.2787 & 0.3924 & 0.500 & 0.460 & 0.440 \\
& HetHunt (+rarity) & -- & -- & -- & -- & \textbf{0.8614} & \textbf{0.8663} & \textbf{1.000} & \textbf{1.000} & \textbf{1.000} \\
\bottomrule
\end{tabular}
\parbox{\textwidth}{%
\centering
\footnotesize
Level 1 is a triage upper bound, not a hunting baseline, for QUT-DV25 separability under engineered summaries.
}
\end{table*}

This subsection addresses \textbf{RQ1} and \textbf{RQ2}.  It first evaluates package-level triage signal under processed non-pivot features and canonical graph representations, and then tests whether HetHunt's edge-derived suspiciousness scores can support analyst-facing package prioritization. Table~\ref{tab:qut_three_level} reports the three-level comparison: processed-feature baselines for package-level classification, GNN backbones over canonical graphs, and the HetHunt suspicious-score ranking pipeline.

The three levels in Table~\ref{tab:qut_three_level} should be interpreted as evaluating different aspects of the problem rather than competing under the same setting.
First, Level~1 establishes a strong upper bound for package-level classification on QUT-DV25. Because these baselines operate on processed behavioral summaries and derived tabular evidence, they have access to a richer feature space than the graph backbones. Their near-perfect results therefore show that QUT-DV25 is highly separable when strong engineered features are available. At the same time, these processed features are aggregated behavioral descriptors rather than direct edge-level hunting pivots, and they do not naturally yield unified, graph-structured evidence that analysts can inspect in the same way as surfaced graph interactions.

Second, Level~2 isolates the contribution of canonical graph representations under lightweight graph-derived node features. These features provide graph-local distinguishability among nodes of the same type without using the original processed QUT-DV25 tabular feature vectors or raw string embeddings. Under this setting, all four graph
backbones learn meaningful package-level decision boundaries. Among them, HGT achieves the strongest overall package-level performance, with the best accuracy, recall, F1, AUROC, and AUPRC, while GAT yields the highest precision. This suggests that canonical structural features are sufficient for heterogeneous GNNs to extract useful triage signal from the graph representation. Nevertheless, Level~2 remains a package-level graph classification baseline and does not itself produce analyst-facing pivots.

\begin{table}[!htbp]
\centering
\scriptsize
\caption{Backbone trade-off on QUT-DV25}
\label{tab:qut_backbone_tradeoff}
\begin{tabular}{lrr}
\toprule
Metric & R-GCN & HGT \\
\midrule
Inference (ms/graph) & \textbf{8.5} & 14.3 \\
\midrule
Cls. AUROC & 0.8124 & \textbf{0.8859} \\
Cls. F1 & 0.6931 & \textbf{0.7955} \\
\midrule
Hunt AUROC & \textbf{0.8614} & 0.8564 \\
Hunt AUPRC & \textbf{0.8663} & 0.8600 \\
Hunt P@100 & 1.000 & 1.000 \\
\bottomrule
\end{tabular}
\end{table}

Table~\ref{tab:qut_backbone_tradeoff} motivates the use of R-GCN in the pivot-producing stage, despite HGT's stronger package-level classification performance. While HGT achieves better package prediction, R-GCN provides stronger rarity-calibrated hunting ranking, lower inference latency, and a simpler relation-indexed interface for reporting edge-level pivots.

Third, Level~3 evaluates the practical hunting utility of the full HetHunt pipeline. In this setting, the goal is no longer threshold-based package classification, but graph-level prioritization using suspicious scores derived from surfaced edge-level evidence. A key finding is that the base suspicious score is not sufficient by itself: without rarity reranking, graph-level ranking remains weak (AUROC$=0.2787$, AUPRC$=0.3924$). However, once edge-level evidence is calibrated against the benign population, ranking quality improves sharply (AUROC$=0.8614$, AUPRC$=0.8663$), and top-of-list precision becomes perfect up to $K=100$.

These results suggest a three-part interpretation of QUT-DV25: processed-feature baselines provide the strongest package-level classifiers, canonical graphs provide a weaker but meaningful relational representation, and HetHunt adds the distinct capability of converting graph explanations into analyst-facing triage rankings.

\subsubsection{Backbone Choice for Hunting}

Although HGT is the strongest Level-2 package classifier, we use R-GCN as the default Level-3 hunting backbone because the final objective is edge-derived suspiciousness ranking rather than package classification alone. As shown in Table~\ref{tab:qut_backbone_tradeoff}, HGT achieves higher classification AUROC and F1, but R-GCN provides slightly stronger rarity-calibrated hunting AUROC/AUPRC, lower inference latency, and a simpler relation-indexed edge interface for mapping explanation scores back to typed runtime pivots. This result highlights that classifier strength and hunting utility are related but not identical objectives: the best package-level backbone is not
necessarily the best operational backbone for analyst-facing pivot generation.

\subsection{OSPTrack as a Noisy Event-Level Stress Test}

This subsection addresses \textbf{RQ2} and \textbf{RQ3} under noisy event-level telemetry. OSPTrack is more challenging than QUT-DV25 because many graphs contain dense package-manager, registry, dependency-installation, and language-runtime activity, while some malicious labels are only weakly reflected in the observed install/import trace. We therefore treat OSPTrack as a stress test rather than a deployment-ready low-FPR alerting benchmark.

Although OSPTrack could be aggregated into coarser QUT-like summaries, doing so would trade pivot specificity for statistical stability. 
We keep OSPTrack at event level because it preserves concrete domains, IPs, paths, and commands as potential IOC-level pivots, allowing us to evaluate the harder observability regime where such artifacts exist but are easily dominated by package-manager, registry, and runtime noise.

\begin{table}[!htbp]
\centering
\caption{OSPTrack graph-level hunting results under noisy event-level telemetry.}
\label{tab:osp-hunting-results}
\footnotesize
\setlength{\tabcolsep}{3pt}
\begin{tabular}{p{1.5cm}ccccc}
\toprule
Method & Scope & AUROC & AUPRC & P@10 & P@100 \\
\midrule
Base suspicious score & -- & 0.5954 & 0.4387 & 0.0 & 0.0 \\
Rarity reranking & NET-only & 0.6413 & 0.4271 & 0.0 & 0.0 \\
Rarity reranking & ALL & 0.0800--0.3184 & 0.2149--0.3084 & 0.0 & 0.0 \\
\bottomrule
\end{tabular}
\end{table}

Table~\ref{tab:osp-hunting-results} shows weak but non-random ranking signal, with poor top-of-list precision. OSPTrack therefore supports candidate edge-level inspection, but not reliable graph-level thresholding.
Rarity calibration on OSPTrack is relation-dependent. Applying rarity to all relation types is unstable because long-tailed FILE and CMD entities amplify benign outliers, while restricting rarity to network-related relations yields the most stable improvement.
Detailed operating-point checks are provided in Appendix~\ref{app:operating-points}. They confirm that QUT-DV25 remains stable across threshold choices, whereas OSPTrack low-FPR thresholding remains unreliable under the evaluated edge-derived score.

\subsection{Explanation and Suspiciousness Baselines}

We next isolate the contribution of each suspiciousness component for \textbf{RQ2} and \textbf{RQ3}. For each
dataset, we compare random ranking, raw graph-level model probability,
rarity-only scoring, raw explainer importance, filtering alone, and the full
HetHunt pipeline with filtering and rarity calibration. This comparison
separates package-level ranking signals from edge-derived hunting scores.

\begin{table}[t]
\centering
\scriptsize
\caption{Package-level triage and edge-derived suspiciousness ranking on QUT-DV25.}
\label{tab:qut_suspiciousness}
\begin{tabular}{p{1.5cm}rrrrr}
\toprule
Method & AUROC & AUPRC & P@10 & P@50 & P@100 \\
\midrule
Random-K & 0.358 & 0.416 & 0.10 & 0.22 & 0.32 \\
Raw model probability & 0.825 & 0.826 & 1.00 & 0.98 & 0.97 \\
Rarity-only & 0.869 & 0.879 & 1.00 & 1.00 & 1.00 \\
HetHunt raw & 0.357 & 0.432 & 0.40 & 0.46 & 0.49 \\
HetHunt + filtering & 0.279 & 0.392 & 0.50 & 0.46 & 0.44 \\
HetHunt + filtering + rarity & \textbf{0.876} & \textbf{0.880} & \textbf{1.00} & \textbf{1.00} & \textbf{1.00} \\
\bottomrule
\end{tabular}
\end{table}

Table~\ref{tab:qut_suspiciousness} reports the QUT-DV25 component ablation. 
The results show that stable aggregated telemetry makes benign rarity a strong prior: rarity-only already achieves high AUROC/AUPRC and perfect top-$K$ precision. The full HetHunt pipeline with filtering and rarity calibration performs slightly better overall, reaching the best AUROC/AUPRC while preserving perfect P@10, P@50, and P@100. 
In contrast, the raw edge-importance score is weak, and filtering without rarity does not by itself produce a strong graph-level ranking signal. 
This is expected because the edge mask is trained for prediction fidelity rather than maliciousness, while filtering is designed to remove environmental noise and improve pivot interpretability rather than to act as a standalone classifier. 
These results support the central separation in HetHunt: model-facing edge importance, benign-population rarity, and analyst-facing suspiciousness are distinct signals, and effective graph-level hunting requires calibrating surfaced evidence against benign behavior.

\begin{table}[t]
\centering
\caption{Package-level triage and edge-derived suspiciousness ranking on OSPTrack. Rarity is restricted to network-related relations for the final HetHunt variant.}
\label{tab:osp-hunting}
\begin{tabular}{p{2cm}ccccc}
\toprule
Method & AUROC & AUPRC & P@10 & P@50 & P@100 \\
\midrule
Random-K & 0.0795 & 0.1852 & 0.00 & 0.00 & 0.00 \\
Raw model probability & \textbf{0.9176} & \textbf{0.7367} & \textbf{0.400} & \textbf{0.580} & \textbf{0.660} \\
Rarity-only & 0.5858 & 0.3899 & 0.00 & 0.00 & 0.28 \\
HetHunt raw & 0.4948 & 0.3176 & 0.00 & 0.10 & 0.16 \\
HetHunt + filtering & 0.6030 & 0.4384 & 0.00 & 0.00 & 0.00 \\
HetHunt + filtering + rarity & 0.6494 & 0.4276 & 0.00 & 0.00 & 0.00 \\
\bottomrule
\end{tabular}
\end{table}

Table~\ref{tab:osp-hunting} reports the same comparison on OSPTrack. Here, raw model probability is the strongest graph-level ranking signal, but it does not provide edge-level pivots. Among edge-derived suspicious scores, the full HetHunt pipeline achieves the best AUROC, improving over rarity-only, raw edge importance, and filtering alone. However, top-of-list precision remains weak for the edge-derived score, indicating that OSPTrack is not suitable for low-FPR alerting using the current edge aggregation alone. We therefore interpret OSPTrack as a noisy event-level stress test rather than a deployment-ready thresholding setting.

\paragraph{Explainer-weight sensitivity}

We also test the sensitivity of the raw edge-mask score to PGExplainer
regularization weights with rarity disabled. Across the tested $\lambda_{\mathrm{sp}}$ and $\lambda_{\mathrm{ent}}$ settings, raw edge-derived ranking remains weak and far below the rarity-calibrated HetHunt result, indicating that tuning the explainer loss alone is
insufficient for strong hunting performance. Full results are reported in Appendix~\ref{app:sensitivity}.

\subsection{Analyst-Usefulness Annotation}
\label{sec:analyst-usefulness}

\begin{table*}[t]
\centering
\caption{Analyst-usefulness annotation of top-5 surfaced pivots. }
\caption*{Results are reported over 50 packages per dataset and three scoring variants.}
\label{tab:analyst-usefulness}
\scriptsize
\begin{tabularx}{\textwidth}{llp{3.1cm}p{3.0cm}X}
\toprule
Dataset & Method & Evidence-level distribution & Dominant relation type & Interpretation \\
\midrule
QUT-DV25 
& Raw explainer 
& 50.8\% TTP / 49.2\% contextual 
& PROC$\rightarrow$NET, PROC$\rightarrow$CMD, PROC$\rightarrow$FILE 
& Aggregated telemetry surfaces TTP/contextual pivots. \\

QUT-DV25 
& Filtering 
& 50.8\% TTP / 49.2\% contextual 
& PROC$\rightarrow$CMD, PROC$\rightarrow$FILE, PROC$\rightarrow$NET 
& Filtering preserves abstraction-bounded hunting evidence. \\

QUT-DV25 
& Filtering + rarity 
& 100.0\% contextual 
& PROC$\rightarrow$WRITE$\rightarrow$FILE, PROC$\rightarrow$READ$\rightarrow$FILE 
& Rarity improves graph ranking, but pivots remain contextual. \\

OSPTrack 
& Raw explainer 
& 68.7\% IOC / 8.1\% contextual / 23.2\% telemetry gap 
& PROC$\rightarrow$READ$\rightarrow$FILE 
& Concrete artifacts exist, but are dominated by package-local file reads. \\

OSPTrack 
& Filtering 
& 68.4\% IOC / 8.5\% contextual / 23.1\% telemetry gap 
& PROC$\rightarrow$READ$\rightarrow$FILE 
& Filtering alone does not remove event-level file-artifact dominance. \\

OSPTrack 
& Filtering + rarity 
& 68.6\% IOC / 8.2\% contextual / 23.3\% telemetry gap 
& PROC$\rightarrow$READ$\rightarrow$FILE 
& Event-level pivots require stronger relation-aware calibration and validation. \\
\bottomrule
\end{tabularx}
\end{table*}

Ranking metrics show whether malicious packages appear near the top of an inspection list, but not whether the surfaced edges are useful pivots. To address \textbf{RQ4}, we annotate the top-5 pivots for 50 packages from each dataset under raw explainer, filtering, and filtering-plus-rarity scoring. Each edge is labeled as IOC-level, TTP-level, contextual, telemetry gap, or missing evidence, following Table~\ref{tab:evidence-levels}. The goal is to characterize the investigation handle HetHunt provides, not to confirm maliciousness.

To reduce annotation bias, the usefulness labels were independently cross-validated by a second annotation pass. The two passes achieved 96.3\% row-level agreement on QUT-DV25 and 97.1\% on OSPTrack, with Cohen's $\kappa$ of 0.919 and 0.939, respectively. Disagreements were concentrated in boundary cases between TTP-level and telemetry-gap labels for syscall-only QUT pivots, and between IOC-level, contextual, and telemetry-gap labels for package-local or environment-driven OSP pivots. Detailed agreement statistics are reported in Appendix~\ref{app:analyst-annotation}.

Table~\ref{tab:analyst-usefulness} summarizes the adjudicated annotation results. The results show that high ranking quality and high pivot specificity are not the same property. On QUT-DV25, raw and filtered outputs split between TTP-level and contextual evidence, while rarity-calibrated outputs become entirely contextual. This reflects QUT-DV25’s aggregated but stable telemetry: it supports strong graph-level suspiciousness ranking, but the surfaced pivots remain bounded by abstract behavioral summaries rather than concrete IOCs. On OSPTrack, most surfaced pivots are IOC-level artifacts, but they are overwhelmingly package-local file-read pivots. This confirms the opposite regime: event-level telemetry preserves concrete artifacts, but the resulting pivots can be dominated by package files, runtime artifacts, and telemetry gaps. These findings support our central interpretation of HetHunt as a telemetry-bounded hunting aid rather than an automatic IOC extractor.

\subsection{Observability Analysis}
\label{sec:observability}

For \textbf{RQ4}, we further examine whether telemetry volume corresponds to useful hunting observability. We compute observability from the canonical graph by excluding
structural \textsc{Load} edges and counting non-\textsc{Load} behavior edges
together with behavior-family richness. QUT-DV25 uses the stored
fixed-threshold buckets from evaluation, while OSPTrack is re-bucketed into
behavior-edge-volume tertiles because the fixed thresholds collapse all OSP
test graphs into the High bucket. The detailed bucket construction and the
QUT-DV25 bucket results are reported in Appendix~\ref{app:observability-buckets}.

Table~\ref{tab:osp-buckets} reports the OSPTrack quantile analysis. The
result shows that telemetry volume is not equivalent to hunting observability:
the highest-volume bucket contains only benign packages, suggesting that dense
event-level traces are dominated by package-manager, dependency-installation,
or runtime activity rather than malicious evidence. The low-volume bucket
achieves stronger ranking quality than the medium bucket, indicating that
cleaner traces can be more useful than larger but noisier traces. This supports our observability-bounded interpretation: hunting utility depends not only on how much telemetry is recorded, but also on whether the recorded entities are specific, stable, and semantically useful.

\begin{table}
\centering
\scriptsize
\caption{OSPTrack observability buckets by behavior-edge volume quantiles.}
\label{tab:osp-buckets}
\begin{tabular}{lrrrrr}
\toprule
Bucket & n & Pos/Neg & Behavior edges & AUROC & AUPRC \\
\midrule
Low & 421 & 225/196 & 695--1721 & 0.837 & 0.792 \\
Medium & 420 & 172/248 & 1742--3447 & 0.517 & 0.534 \\
High & 420 & 0/420 & 3468--150039 & -- & -- \\
\bottomrule
\end{tabular}
\end{table}

\subsection{Hunting Pivot Case Studies (RQ5)}
\label{sec:reliability-analysis}

\begin{table}[t]
\centering
\scriptsize
\caption{Telemetry-Bounded Hunt Outputs}
\label{tab:case-studies}
\begin{tabular}{p{1.8cm}p{0.4cm}p{1.8cm}p{3.0cm}}
\toprule
Case & Dataset & Evidence level & Main point \\
\midrule
10Cent10/10Cent11
& QUT
& Partial / TTP-level
& Aggregated telemetry surfaces behavioral evidence. \\

capmonstercloud* or ligitgays
& OSP
& Contextual / telemetry gap
& Package family recovered, but payload IOC is absent. \\

commonmarker \_pluggable
& OSP
& Candidate IOC-level
& Concrete IP/command pivots are surfaced, but require validation. \\
\bottomrule
\end{tabular}
\end{table}

Finally, we examine representative cases to illustrate how HetHunt's outputs should be interpreted under different telemetry regimes. Table~\ref{tab:case-studies} summarizes three types of telemetry-bounded hunting outcomes, and Appendix~\ref{tab:hunting_mapping}
provides the full package-level mapping. These cases are not intended to treat top-ranked edges as confirmed IOCs; instead, they show what kind of investigation handle is available when the relevant behavior is preserved, abstracted, or absent from the runtime trace.

The cases reinforce the main quantitative result: QUT-DV25 provides stable but abstract TTP-level or contextual evidence, while OSPTrack can surface concrete IOC-level candidates only when those artifacts are recorded and not dominated by environment-driven runtime activity.

\section{Discussion}

\subsection{Observability and Hunting Utility}

Our results show that graph compatibility and hunting utility are different properties. A shared canonical schema can represent both aggregated and event-level telemetry, but the resulting pivots differ in specificity, stability, and operational value. This means that
runtime hunting systems should not be evaluated only by package-level classification accuracy. They should also be assessed by whether the surfaced interactions are specific enough to guide inspection and whether their scores remain reliable under the telemetry noise of the target deployment.

\subsection{Limitations and Threats to Validity}

\textbf{Telemetry-dependent evidence.}
HetHunt can only surface evidence preserved in the runtime trace. If the sandbox does not trigger the malicious path, or if the monitoring pipeline abstracts away concrete artifacts, the explainer cannot recover the missing IOC. Thus, missing concrete indicators should be interpreted as telemetry gaps or untriggered behavior rather than necessarily as explanation failures.

\textbf{No automatic IOC confirmation.}
Surfaced edges are investigation pivots, not confirmed IOCs. Some candidate network outputs may correspond to common infrastructure, registry access, or sandbox-driven behavior. We therefore treat a surfaced edge as a direct match only when it preserves the same artifact reported by external intelligence; otherwise, it is partial, contextual, or candidate evidence.

\textbf{Pooling and rare-behavior dilution.}
HetHunt uses mean pooling over node embeddings to obtain a package-level graph representation. This readout is simple and stable, but it may dilute rare security-relevant behavior in large graphs; for example, thousands of benign file reads may dominate the representation while a single suspicious network connection contributes weakly. Future variants could explore top-$k$ pooling, relation-aware pooling, attention-based readout, or suspiciousness-guided pooling to better preserve rare high-risk interactions.

\textbf{Dependence on rarity calibration.}
Rarity calibration depends on the benign reference population. Incomplete, biased, or environment-mismatched benign data can inflate benign outliers or suppress suspicious behaviors that are common in the reference set. Our results also suggest that strong hunting rankings may rely substantially on benign rarity calibration, especially on QUT-DV25 where rarity-only is already strong. Future ablations should more systematically quantify when the learned graph model and explainer add value beyond rarity-based scoring alone.

\section{Conclusion}

Runtime supply-chain hunting should not be framed solely as package-level detection.  In practice, analysts need telemetry-bounded runtime pivots that explain where to inspect and why a package appears suspicious. 
HetHunt addresses this need by mapping heterogeneous package-execution telemetry into canonical evidence graphs, learning package risk from relational graph structure, and converting edge importance into analyst-facing suspiciousness through filtering and relation-aware benign rarity.

Our evaluation shows that hunting utility depends critically on telemetry granularity and stability. Stable aggregated telemetry can support strong TTP-level triage, whereas event-level telemetry can preserve IOC candidates but also introduces noise that undermines reliable low-FPR alerting. Overall, HetHunt demonstrates that heterogeneous runtime graphs can bridge package-level risk prediction and analyst investigation, provided that explanations are calibrated to the reliability and investigative value of the observed relations.

\section*{Acknowledgments}
This work was supported by industrial funding from JUMPSEC Ltd.


\appendices

\section{Inference Latency}
\label{app:latency}

\textbf{Training protocol.} HetHunt follows a train-then-explain protocol: we use a train-then-explain protocol, train separate encoders for each dataset, and serialize per-package PyG graphs with typed edge lists.

\textbf{Dataset-specific encoders.} Because OSPTrack and QUT-DV25 differ in
telemetry granularity and label distribution, we train separate encoders while
sharing the canonical ontology, graph construction pipeline, and edge-importance
architecture. This preserves dataset-specific behavior distributions while
keeping pivot outputs comparable.

\begin{table}[!htbp]
\centering
\scriptsize
\caption{Inference latency summary. All times are in ms; N/E denotes mean nodes/edges per graph.}
\label{tab:latency_summary}
\begin{tabular}{@{}lrrrrrr@{}}
\toprule
Dataset & \#G & Mean & P50 & P95 & Max & N/E \\
\midrule
QUT-DV25 & 2570 & 301.49 & 215.66 & 708.69 & 4564.63 & 70/81 \\
OSPTrack & 1261 & 5671.93 & 5531.42 & 6471.60 & 10745.52 & 3644/6831 \\
\bottomrule
\end{tabular}
\end{table}

\textbf{Graph handling.} The construction pipeline serializes per-package
graphs, streams them through a PyG DataLoader with mini-batching, and supports
parallel CPU preprocessing. Homogeneous conversion yields one typed edge list
per graph, avoiding per-relation tensor management in the main pipeline. Nodes
use learnable type embeddings rather than token or sentence-transformer
features, reducing preprocessing and memory cost.

Table~\ref{tab:latency_summary} shows that inference cost is dominated by graph
size and explanation workload. QUT-DV25 graphs are small and processed within
sub-second latency for most samples, whereas OSPTrack graphs are much larger
and require more explanation time.

\textbf{Backbone-level efficiency}
In addition to end-to-end inference latency, we measure package-level backbone inference latency on QUT-DV25 to contextualize the choice of the hunting backbone. Table~\ref{tab:qut_backbone_efficiency} summarizes inference time, parameter count, and package-classification performance for the four graph backbones.

\begin{table}[!htbp]
\centering
\scriptsize
\caption{QUT-DV25 backbone trade-off.}
\label{tab:qut_backbone_efficiency}
\begin{tabular}{lrrrr}
\toprule
Backbone & Infer. ms & Params & Cls. AUROC & Cls. F1 \\
\midrule
R-GCN & 8.5  & 81.0K & 0.8124 & 0.6931 \\
GAT   & 4.5  & 49.9K & 0.8199 & 0.6935 \\
HAN   & 7.4  & 34.9K & 0.8554 & 0.7454 \\
HGT   & 14.3 & 4.3K  & 0.8859 & 0.7955 \\
\bottomrule
\end{tabular}
\end{table}
\FloatBarrier

\section{Operating-Point Analysis}
\label{app:operating-points}

\begin{table}[!htbp]
\centering
\caption{Calibration sensitivity of representative RF operating points on QUT-DV25.}
\label{tab:qut_operating_points}
\scriptsize
\setlength{\tabcolsep}{3.5pt}
\begin{tabular}{llrrrrrr}
\toprule
Calib. & Op. point & Thr. & Prec. & Rec. & F1 & FP & FN \\
\midrule
None  & 0.5        & 0.50 & 0.9892 & 0.9923 & 0.9908 & 14 & 10 \\
None  & FPR$\le$.01 & 0.54 & 0.9915 & 0.9892 & 0.9903 & 11 & 14 \\
Platt & 0.5        & 0.50 & 0.9892 & 0.9915 & 0.9904 & 14 & 11 \\
Platt & FPR$\le$.01 & 0.57 & 0.9915 & 0.9892 & 0.9903 & 11 & 14 \\
Iso.  & 0.5        & 0.50 & 0.9818 & 0.9992 & 0.9904 & 24 & 1 \\
Iso.  & FPR$\le$.01 & 0.90 & 0.9915 & 0.9892 & 0.9903 & 11 & 14 \\
\bottomrule
\end{tabular}
\end{table}

\begin{table}[t]
\centering
\caption{OSPTrack operating-point sanity check using graph hunting suspicious scores.}
\label{tab:osp_operating_points}
\scriptsize
\setlength{\tabcolsep}{3pt}
\begin{tabular}{llrrrrrr}
\toprule
Method & Op. point & Thr. & Prec. & Rec. & F1 & FP & FN \\
\midrule
Base & FPR$\le$.01 & -0.0024 & 0.0 & 0.0 & 0.0 & 10 & 397 \\
Rarity & FPR$\le$.01 & -0.0024 & 0.0 & 0.0 & 0.0 & 10 & 397 \\
Base & FPR$\le$.05 & -0.0051 & 0.0 & 0.0 & 0.0 & 38 & 397 \\
Rarity & FPR$\le$.05 & -0.0051 & 0.0 & 0.0 & 0.0 & 38 & 397 \\
Base & FPR$\le$.10 & -0.0094 & 0.0 & 0.0 & 0.0 & 86 & 397 \\
Rarity & FPR$\le$.10 & -0.0094 & 0.0 & 0.0 & 0.0 & 86 & 397 \\
\bottomrule
\end{tabular}
\vspace{1pt}
\begin{minipage}{\columnwidth}
\scriptsize \textit{Note:} Rarity is applied only to NET relations
\end{minipage}
\end{table}

We evaluate deployment-oriented operating points by selecting thresholds on the validation split and testing them on the test split. Table~\ref{tab:qut_operating_points} shows that the QUT-DV25 RF baseline is stable across thresholding and calibration strategies; calibration changes probability scales but has little effect on precision--recall trade-offs. In contrast, Table~\ref{tab:osp_operating_points} shows that OSPTrack edge-derived suspicious scores remain unreliable for low-FPR graph-level alerting, supporting its role as a noisy event-level stress test.

\section{Analyst-Usefulness Annotation Details}
\label{app:analyst-annotation}
For each dataset, we sample 50 packages and inspect the top-5 pivots under raw explainer, filtering, and filtering plus rarity. Each pivot is assigned an evidence label from Table~\ref{tab:evidence-levels}. The annotation characterizes the investigation handle surfaced by HetHunt, not whether the pivot is malicious. Table~\ref{tab:annotation-distribution} reports the full distribution.

\begin{table}[!htbp]
\centering
\caption{Full analyst-usefulness label distribution}
\label{tab:annotation-distribution}
\scriptsize
\begin{tabularx}{\columnwidth}{llrrrrr}
\toprule
Dataset & Method & IOC & TTP & \makecell{Cont.} & \makecell{Tele.\\gap} & Miss \\
\midrule
QUT & Raw & 0 & 127 & 123 & 0 & 0 \\
QUT & Filtering & 0 & 127 & 123 & 0 & 0 \\
QUT & Filtering + rarity & 0 & 0 & 250 & 0 & 0 \\
OSP & Raw & 170 & 0 & 21 & 59 & 0 \\
OSP & Filtering & 171 & 0 & 21 & 58 & 0 \\
OSP & Filtering + rarity & 170 & 0 & 20 & 60 & 0 \\
\bottomrule
\end{tabularx}
\end{table}

\begin{table}[!htbp]
\centering
\scriptsize
\caption{Cross-validation agreement for analyst-usefulness annotation.}
\label{tab:annotation_agreement}
\begin{tabular}{lccc}
\toprule
Dataset & Rows & Agreement & Cohen's $\kappa$ \\
\midrule
QUT-DV25 & 750 & 96.3\% & 0.919 \\
OSPTrack & 750 & 97.1\% & 0.939 \\
\bottomrule
\end{tabular}
\end{table}

To reduce annotation bias, we independently validated the usefulness labels with a second annotation pass and report two agreement measures. Row-level agreement is the fraction of surfaced pivots for which the two annotation passes assign the same evidence label:
\[
\mathrm{Agreement} = \frac{1}{N}\sum_{i=1}^{N}\mathbf{1}[y_i^{(1)} = y_i^{(2)}].
\]
We also report Cohen's $\kappa$~\cite{cohen1960}, which adjusts the observed agreement $p_o$ by the agreement expected by chance $p_e$, as $\kappa=(p_o-p_e)/(1-p_e)$. Here, $p_o$ is the row-level agreement between the two annotation passes, and $p_e$ is computed from the empirical label marginals as $p_e=\sum_{c\in\mathcal{C}} p_c^{(1)}p_c^{(2)}$, where $\mathcal{C}$ is the set of evidence labels and $p_c^{(j)}$ denotes the fraction of pivots assigned label $c$ in annotation pass $j \in \{1,2\}$.

\FloatBarrier

\section{Sensitivity Study}
\label{app:sensitivity}

\begin{figure*}[!htbp]
\centering 
\footnotesize 
\captionsetup{justification=centering} 
\begin{minipage}[t]{0.85\textwidth} \centering \includegraphics[width=\linewidth]{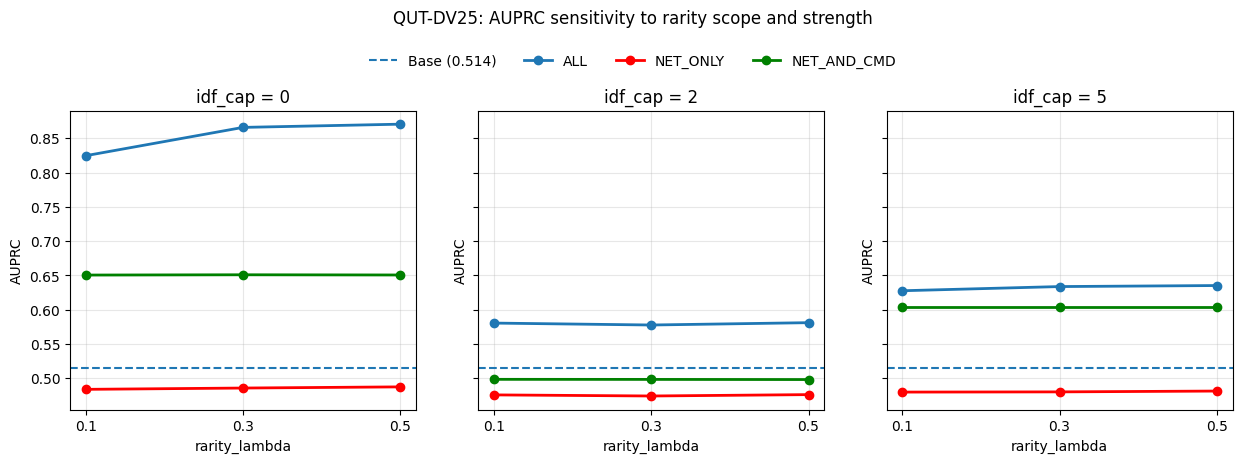} \centerline{(a) QUT-DV25} 
\end{minipage} 
\vspace{0.8em} 
\begin{minipage}[t]{0.85\textwidth} \centering \includegraphics[width=\linewidth]{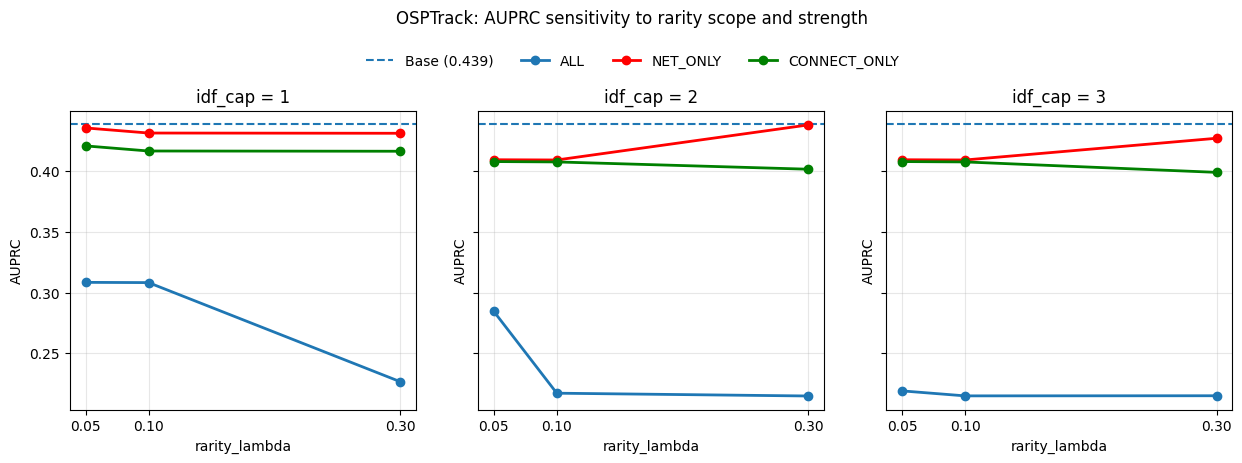} \centerline{(b) OSPTrack} 
\end{minipage} \caption{Sensitivity study of AUPRC to rarity scope and strength.} \label{fig:auprc-sensitivity-study} 
\end{figure*}

\textbf{Rarity scope and strength.}
Figure~\ref{fig:auprc-sensitivity-study} shows that rarity calibration
is more sensitive to relation scope and IDF capping than to the exact
rarity multiplier. On QUT-DV25, the broad ALL scope performs best
when uncapped, consistent with the dataset's stable aggregated
entities, but capped or network-only settings reduce the benefit. On
OSPTrack, applying rarity to all relations is harmful because long-tailed
FILE and CMD entities amplify benign outliers; NET-only rarity gives
the strongest AUPRC, while CONNECT-only rarity is stable but slightly
weaker. These results support relation-aware rarity calibration rather
than uniformly applying rarity to all edge types.

\section{Graph Representation and Explainer Implementation}
\label{app:graph-implementation}

HetHunt constructs heterogeneous runtime graphs using the canonical ontology in
Table~\ref{tab:ontology}, which defines $T_n=6$ node types and $T_e=11$ edge
types. The OSP and QUT columns show which types are instantiated in each dataset;
unobserved types yield no instances. For implementation, each graph is converted
into a typed edge-list, with node and edge types preserved through $\phi$ and
$\psi$ for R-GCN encoding and pivot reporting.

\textbf{Dataset-specific parsing.}
Each parser emits canonical edge events with the shared schema:
(source type, source key, relation, target type, target key, attributes). For each
package, we create a \texttt{PKG$\!\to\!$PROC (LOAD)} edge for the installation
or import process and map observed behaviors to
\texttt{PROC}-rooted interactions, yielding the common two-hop structure
\texttt{PKG$\!\to\!$PROC$\!\to\!$\{FILE, NET, CMD, SYSCALL\}}.
OSPTrack maps structured runtime logs to canonical relations: file records to
\texttt{\seqsplit{PROC$\!\to\!$FILE(READ/WRITE/DELETE)}}, DNS records to
\texttt{\seqsplit{PROC$\!\to\!$NET(DNS\_QUERY/RESOLVE)}}, socket endpoints to
\texttt{PROC$\!\to\!$NET(CONNECT)}, and command strings to
\texttt{PROC$\!\to\!$CMD(EXEC)}. Raw strings are normalized to reduce noise from
ephemeral temporary paths and platform-specific prefixes. QUT-DV25 joins
dependency, syscall, file-access, TCP, and pattern-trace tables by package
identifier, producing \texttt{PKG$\!\to\!$PKG(DEPEND)},
\texttt{PROC$\!\to\!$SYSCALL(INVOKE)},
\texttt{PROC$\!\to\!$FILE(READ/WRITE)},
\texttt{PROC$\!\to\!$NET(CONNECT)}, and
\texttt{PROC$\!\to\!$CMD(EXEC)} edges.

\textbf{Typed edge-list conversion.}
Given canonical edge events, each package is represented as
$G=(V,E,\phi,\psi)$, where $\phi:V\to\{1,\dots,T_n\}$ assigns node types and
$\psi:E\to\{1,\dots,T_e\}$ assigns relation types. Each node $v$ is initialized
with a learnable type embedding $\mathbf{x}_v=\mathbf{W}_{\phi(v)}$, where
$\mathbf{W}\in\mathbb{R}^{T_n\times d}$. This representation lets R-GCN apply
relation-specific transformations conditioned on $\psi(e)$ and lets the
edge-mask module score one indexable edge set. We evaluate HAN and HGT as
heterogeneous classification baselines, but use the R-GCN typed edge-list view
for the unified hunting pipeline so that top-$K$ edge scores map back to
human-readable relations and entities.

\begin{table}[t]
\centering
\scriptsize
\caption{Sensitivity of QUT-DV25 edge-derived ranking to PGExplainer regularization weights. Rarity calibration is disabled or neutralized in this sweep to isolate the learned edge-mask score.}
\label{tab:qut_pgexp_sensitivity}
\begin{tabular}{rrrrr}
\toprule
$\lambda_{sp}$ & $\lambda_{ent}$ & AUROC & AUPRC & F1@0.10 \\
\midrule
0.1  & 0    & 0.214 & 0.352 & 0.021 \\
0.1  & 0.01 & 0.211 & 0.350 & 0.004 \\
0.1  & 0.1  & 0.216 & 0.352 & 0.008 \\
1.0  & 0    & 0.233 & 0.358 & 0.024 \\
1.0  & 0.01 & 0.253 & 0.380 & -- \\
1.0  & 0.1  & 0.213 & 0.351 & 0.004 \\
10.0 & 0    & 0.230 & 0.359 & 0.028 \\
10.0 & 0.01 & 0.214 & 0.353 & 0.024 \\
10.0 & 0.1  & \textbf{0.302} & \textbf{0.404} & -- \\
\bottomrule
\end{tabular}
\end{table}

\textbf{Explainer-loss weights.}
We next evaluate the sensitivity of the edge-mask training objective in Eq.~\ref{eq:edge_mask_training_objective}. 
We keep the trained R-GCN backbone fixed, use the QUT-DV25 test split, and vary 
$\lambda_{\mathrm{sp}}\in\{0.1,1.0,10.0\}$ and 
$\lambda_{\mathrm{ent}}\in\{0,0.01,0.1\}$. 
The target keep ratio is fixed at 0.05. 
To isolate the learned edge-mask score from post-hoc calibration effects, rarity weighting is disabled or neutralized while the same filtering configuration is retained.

Table~\ref{tab:qut_pgexp_sensitivity} shows that the learned edge-mask score is sensitive to the PGExplainer regularization weights, but remains weak without rarity calibration. 
The best setting in this sweep is $\lambda_{\mathrm{sp}}=10.0$ and $\lambda_{\mathrm{ent}}=0.1$, reaching 0.302 AUROC and 0.404 AUPRC. 
The default setting $(\lambda_{\mathrm{sp}}=1.0,\lambda_{\mathrm{ent}}=0.01)$ is not the strongest raw edge-ranking configuration in this sweep. 
However, all tested settings remain far below the rarity-calibrated HetHunt result reported in the main evaluation, reinforcing that model-facing edge importance alone is not a reliable suspiciousness signal and that benign rarity calibration is the key step for converting surfaced edges into analyst-facing rankings.

\subsection*{Observability Bucket Construction}
\label{app:observability-buckets}

We compute observability buckets from the canonical graph used during explainer
evaluation. For each graph, structural \textsc{Load} edges are excluded, and
all remaining edges are counted as behavior edges. Each behavior edge is mapped
to a coarse family---\textsc{Net}, \textsc{Cmd}, \textsc{File},
\textsc{Syscall}, \textsc{Proc}, or \textsc{Other}---based on its relation
name or destination node type. Let $b(G)$ denote the number of behavior edges
and $r(G)$ the number of behavior families present in graph $G$. The stored
observability bucket is assigned as:
\[
\text{bucket}(G)=
\begin{cases}
\textsc{LoadOnly}, & b(G) \le 0,\\
\textsc{Low}, & b(G) < 10 \text{ or } r(G) \le 1,\\
\textsc{Medium}, & b(G) < 50 \text{ or } r(G) \le 2,\\
\textsc{High}, & \text{otherwise}.
\end{cases}
\]

\begin{table*}[!htbp]
\centering
\scriptsize
\caption{Full telemetry-bounded hunting outputs. Report-backed rows are
compared with public reports when available; Candidate rows illustrate
telemetry-preserved pivots that require analyst validation.}
\label{tab:hunting_mapping}
\begin{tabularx}{\textwidth}{p{2cm} p{1.2cm} X X p{1cm} X}
\toprule
Package & Report Source & Reported Behavior & Surfaced Hunting Evidence & Mapping & Predicted Suspicious Output \\
\midrule
\texttt{10Cent10} (QUT) 
& JFrog~\cite{polkovnychenko2021pythonmalware}
& Connectback shell to hardcoded address \texttt{104.248.19.57}. 
& Rarity-adjusted explanation promotes install-phase network and command/syscall-pattern evidence. 
& Partial 
& \texttt{PROC|CONNECT|NET}: \texttt{install} $\rightarrow$ \texttt{NET:ip::62}; \texttt{PROC|CONNECT|NET}: \texttt{install} $\rightarrow$ \texttt{NET:port::21}. \\

\texttt{10Cent11} (QUT) 
& JFrog~\cite{polkovnychenko2021pythonmalware} 
& Same campaign as \texttt{10Cent10}; connectback shell behavior to \texttt{104.248.19.57}. 
& Explanation surfaces install-phase socket/network-related and command-pattern evidence. 
& Partial 
& \texttt{PROC|CONNECT|NET}: \texttt{install} $\rightarrow$ \texttt{NET:ip::51}; \texttt{PROC|CONNECT|NET}: \texttt{install} $\rightarrow$ \texttt{NET:port::18}. \\

\texttt{capmonstercloud*} \newline typosquats (OSP) 
& Veracode~\cite{veracode2024typosquatting}
& Typosquatted variants of \texttt{capmonstercloudclient}; reports describe malicious install hooks and zgRAT/data-stealer payloads. 
& The case study recovers the reported typosquatted package family and its install-time execution context, but the surfaced top edges are dominated by package-manager invocation and package file access. 
& Package-level 
& \texttt{PROC|EXEC|CMD}: \texttt{install} $\rightarrow$ package-manager invocation; no payload URL, C2 endpoint, or webhook edge is surfaced. \\

\texttt{ligitgays} (OSP) 
& Unit42~\cite{benhai2023sixmalicious}
& Remote payload retrieval and W4SP-like credential/crypto-wallet theft behavior. 
& The case study recovers the reported package and install-time execution context, but does not surface the reported remote payload URL or webhook. 
& Package-level 
& \texttt{PROC|EXEC|CMD}: \texttt{install} $\rightarrow$ package-manager invocation; no remote payload IOC is surfaced. \\

\texttt{bettercolors} (OSP) 
& PyPI \newline malicious-package~\cite{lakshmanan2024pypihalts} 
& Color/colorama-themed malicious package activity and obfuscated install-time payloads. 
& The explanation surfaces installation workflow and package file-access evidence, but not an independently verifiable malicious payload or network IOC. 
& Package-level 
& \texttt{PROC|EXEC|CMD}: \texttt{install} $\rightarrow$ package-manager invocation; evidence remains workflow-level rather than payload-level. \\

\texttt{commonmarker\newline\_pluggable} (OSP) 
& -- 
& No public report used; included as a telemetry-resolution candidate. 
& After registry-domain filtering, rarity-adjusted explanations surface concrete destination IPs and RubyGems install commands. 
& Candidate 
& \texttt{PROC|CONNECT|NET}: \texttt{install} $\rightarrow$ \texttt{NET:ip::8.8.4.4}; \texttt{PROC|CONNECT|NET}: \texttt{install} $\rightarrow$ \texttt{NET:ip::151.101.*.227}. \\

\texttt{ach-client} (OSP) 
& -- 
& No public report used; included as a telemetry-resolution candidate. 
& Rarity-adjusted explanations surface concrete destination IPs and RubyGems install commands after filtering common registry domains. 
& Candidate 
& \texttt{PROC|CONNECT|NET}: \texttt{install} $\rightarrow$ \texttt{NET:ip::151.101.*.227}; \texttt{PROC|EXEC|CMD}: \texttt{install} $\rightarrow$ RubyGems install command. \\

\texttt{fluent\_plugin\newline-cloudwatch-logs\newline-foxtrot9} (OSP) 
& -- 
& No public report used; included as a telemetry-resolution candidate. 
& Rarity-adjusted explanations surface install-time RubyGems commands and concrete destination IPs. 
& Candidate 
& \texttt{PROC|CONNECT|NET}: \texttt{install} $\rightarrow$ \texttt{NET:ip::151.101.*.227}; \texttt{PROC|EXEC|CMD}: \texttt{install} $\rightarrow$ RubyGems install command. \\
\bottomrule
\end{tabularx}
\end{table*}

\begin{table}[!htbp]
\centering
\scriptsize
\caption{QUT-DV25 observability buckets using fixed-threshold graph observability labels.}
\label{tab:qut-observability-app}
\begin{tabular}{lrrrrr}
\toprule
Bucket & n & Pos/Neg & Beh. edges  & AUROC & AUPRC \\
\midrule
Low    & 1    & 1/0       & 9.0   & --    & --    \\
Medium & 129  & 113/16    & 30.7  & 0.828 & 0.974 \\
High   & 2440 & 1182/1258 & 77.4 & 0.868 & 0.862 \\
\bottomrule
\end{tabular}
\end{table}
\FloatBarrier

QUT-DV25 reporting uses these stored fixed-threshold buckets. In contrast,
the fixed-threshold rule is not informative for OSPTrack because all held-out
OSPTrack graphs exceed the \textsc{High} threshold. We therefore use a
post-hoc quantile re-bucketing for the OSPTrack observability analysis in
Section~\ref{sec:observability}: graphs with $b(G)>0$ are split into
\textsc{Low}, \textsc{Medium}, and \textsc{High} quantiles by behavior-edge count using the 33rd and 66th percentiles on the test set. Graphs with
$b(G)=0$ remain \textsc{LoadOnly}. This quantile mode makes the OSPTrack
analysis dataset-relative and avoids the degenerate single-bucket summary
produced by the fixed thresholds.
As shown in Table~\ref{tab:qut-observability-app}, QUT-DV25 is dominated by the \textsc{High} bucket under the fixed-threshold rule, reflecting the availability of stable aggregated behavioral summaries for most samples. The \textsc{Low} bucket contains only one package and is not meaningful for ranking, while the \textsc{Medium} and \textsc{High} buckets both show strong ranking quality.

\section{Full Case Studies}
\label{app:case-studies}

This appendix expands the case-study summary in
Section~\ref{sec:reliability-analysis}. We use these cases to characterize the
analyst-facing evidence surfaced by HetHunt, not to treat top-ranked edges as
confirmed IOCs. Report-backed rows are compared with public reports when
available, while candidate rows are telemetry-preserved leads that require
analyst validation.

The cases illustrate the observability-bounded nature of HetHunt's outputs:
QUT-DV25 often localizes behavior only to aggregated install-phase network or
command/syscall patterns, whereas OSPTrack can preserve concrete event-level
artifacts but is more easily dominated by package-manager, registry, pseudo-file,
and runtime activity. Missing payload URLs or C2 endpoints are therefore
interpreted as telemetry gaps or untriggered behavior rather than explanation
failures. Table~\ref{tab:hunting_mapping} reports the full mapping.

\end{document}